# Expanding the trilayer Ruddlesden-Popper nickelate family: Synthesis and characterization of $Sm_4Ni_3O_{10-\delta}$ single crystals

*Yuhang Zhang, Tian-Yi Li, Xiyu Zhu*, Ying-Jie Zhang, Shengtai Fan, Qing Li* and Hai-Hu Wen**

National Laboratory of Solid State Microstructures and Department of Physics, Center for Superconducting Physics and Materials, Collaborative Innovation Center of Advanced Microstructures, Nanjing University, Nanjing 210093, China
*Corresponding author: zhuxiyu@nju.edu.cn; liqing1118@nju.edu.cn; hhwen@nju.edu.cn;

**Abstract:** The discovery of high-temperature superconductivity in Ruddlesden-Popper (RP) nickelates has attracted significant attention. Bulk superconductivity emerges under pressure in trilayer nickelates $La_4Ni_3O_{10-\delta}$ ($T_c \approx 30$ K) and $Pr_4Ni_3O_{10-\delta}$ ($T_c \approx 40.5$ K), where the reduced ionic radius of $Pr^{3+}$ may generate internal chemical pressure and enhance $T_c$. However, synthesizing trilayer RP phases with smaller rare-earth elements (*Ln*) is extremely challenging. So far, only the La, Pr, and Nd analogues have been synthesized with stable phases in the single rare-earth form. Here we report the first successful high-pressure and high-temperature (HPHT) synthesis of samarium-based compound $Sm_4Ni_3O_{10-\delta}$. Magnetization and transport measurements consistently confirm a density wave (DW) transition at ∼180 K at ambient pressure. Through a careful fitting to the structural data of $Sm_4Ni_3O_{10-\delta}$, it is found that the bond angle of (Ni−O−Ni) associating with the interlayer apical oxygen is much smaller than 180°, which was assumed to be the key factor for the occurrence of superconductivity. By applying pressures up to 80 GPa, despite partial suppression of insulating behavior and the DW order, but superconductivity is not observed in our present study. Density functional theory calculations suggest that the $\boldsymbol{3d_{z^2}}$ and $\boldsymbol{3d_{x^2-y^2}}$ are separated from other $t_{2g}$ orbitals and make a primary contribution to the Fermi surface. The newly synthesized trilayer nickelate $Sm_4Ni_3O_{10-\delta}$ offers a unique platform for probing the fundamental physics of RP nickelates.

## 1. Introduction

The discovery of high temperature superconductivity in the nickelate systems has put them as the third major class of unconventional high-temperature superconductors after cuprates and iron-based compounds, which is very important for unraveling the high-

$T_c$ superconductivity mechanism [1-4]. Up to date, superconductivity has been discovered in several nickelate systems, among them there exist two major groups: (1) Infinite-layer or reduced RP phases, epitomized by $Nd_{1-x}Sr_xNiO_2$, $Nd_6Ni_5O_{12}$ and $Sm_{0.79}Eu_{0.12}Ca_{0.04}Sr_{0.05}NiO_2$ thin films, where $Ni^{1+}$ ($3d^9$) mimics the electronic configuration of cuprate $Cu^{2+}$ [3,5-9]; (2) RP nickelates ($Ln_{n+1}Ni_nO_{3n+1}$, n=2 and 3) exhibiting superconductivity under high-pressure condition. In 2023, an exciting breakthrough occurred for the discovery of high-temperature superconductivity at about 80 K in $La_3Ni_2O_7$ with a cationic state of $Ni^{+2.5}$ ($3d^{7.5}$), where the $d_{z^2}$ band forms $c$-axis bonding/antibonding states in Ni−O−Ni, splitting the $d_{z^2}$ level into occupied and unoccupied states [4, 10-13]. Although multiple predictions suggest that $Ln_3Ni_2O_7$ may exhibit superconductivity with potentially even higher $T_c$, the synthesis of single rare-earth $Ln_3Ni_2O_7$ remains confined to $Ln$ = La up to now [14-16]. Inspiringly, partial doping of $La_3Ni_2O_7$ with small-radius rare-earth elements leads to a dramatic enhancement of superconductivity. For instance, bulk superconductivity emerges in $La_2PrNi_2O_7$, whereas $T_c$ exceeds 90 K in $La_2SmNi_2O_7$ [17-18]. Besides, two polymorphic stacking sequences, conventional LNO-2222 and the newly discovered LNO-1313, have been identified in $La_3Ni_2O_7$ [19-21]. Most recently, ambient-pressure superconductivity was observed in $La_3Ni_2O_7$, $(La,Pr)_3Ni_2O_7$ and $(La,Sr)_3Ni_2O_7$ thin films [22-25], enabling the investigation of physical properties without external pressure [26-29]. Concurrently, high-pressure superconductivity has also been discovered in the hybrid nickelate $La_5Ni_3O_{11}$, which adopts the distinctive LNO-1212 stacking configuration [30,31].

Following the breakthrough discovery of superconductivity in bilayer nickelates, trilayer analogues $Ln_4Ni_3O_{10-\delta}$ ($Ln$ = La, Pr, Nd) have garnered significant research interest. In previous reports, neutron scattering and X-ray data have revealed an intertwinement of spin-density-wave (SDW) and charge-density-wave (CDW) at 136 K in $La_4Ni_3O_{10-\delta}$ [32], whereas at 157 K in $Pr_4Ni_3O_{10-\delta}$ [33,34]. Similar DW transition was also observed at 162 K in $Nd_4Ni_3O_{10-\delta}$ [35,36]. In addition, both bilayer and trilayer systems display distorted $NiO_6$ octahedra at ambient pressure [37,38]. Application of high pressure suppresses these structural distortions and DW states, concomitant with the emergence of superconductivity [39-45]. Bulk superconductivity emerges in $La_4Ni_3O_{10-\delta}$ single crystals with $T_c$ of 30 K at 69.0 GPa [42], while $Pr_4Ni_3O_{10-\delta}$ exhibits enhanced transition temperature reaching 40.5 K at 80.1 GPa [45]. The reduced ionic radius of $Pr^{3+}$ simultaneously enhances $T_c$ while necessitating higher applied pressures to achieve superconductivity. Notably, $La_4Ni_3O_{10-\delta}$ with tetragonal structure at ambient pressure, which lacks both DW transitions and superconductivity, demonstrates that suppression of the density-wave state may be prerequisite for the emergence of superconductivity in nickelates [46].

In RP nickelates, using of rare-earth elements with smaller ionic radii may reduce the structural stability as the tolerance factor deviates further from 1, making synthesis more challenging [47, 48]. Consequently, only $La_3Ni_2O_7$ and $Ln_4Ni_3O_{10-\delta}$ ($Ln$ = La, Pr, Nd) have been reported as bilayer and trilayer RP phases with single rare-earth composition [14, 47]. Crucially, partial or even full substitution with smaller rare-earth ions introduces chemical pressure that significantly enhances the superconducting volume and transition temperatures [18, 45]. This underscores the imperative to investigate RP nickelates incorporating smaller rare-earth constituents. In this paper, we report the successful HPHT (3.25 GPa, 1400°C) synthesis of $Sm_4Ni_3O_{10-\delta}$ single crystals, a pioneering trilayer RP nickelate with the smallest rare-earth ion to date. Structural analysis reveals that although $Sm_4Ni_3O_{10-\delta}$ exhibits higher space-group symmetry (*Pbca*) compared to other trilayer nickelates ($P2_1/a$) [38, 42, 45, 49], its Ni−O−Ni bond angle associating with the interlayer apical oxygen atoms deviates severely from 180°. It is argued that this significant deviation hinders the emergence of superconductivity even with pressures up to 80 GPa, although pressure suppresses both insulating behavior and the density-wave order. Our density functional theory calculations on $Sm_4Ni_3O_{10-\delta}$ show a band structure similar to that of $La_4Ni_3O_{10-\delta}$, where the $e_g$ orbitals account for a large proportion around the Fermi level, and the $t_{2g}$ orbitals are away from the Fermi energy. Our results on the new compound $Sm_4Ni_3O_{10-\delta}$ shed new light in understanding the physics and the superconductivity in the RP phases of nickelate family.

## 2. Results and discussion

The $Sm_4Ni_3O_{10-\delta}$ single crystals were grown by using the flux method with a HPHT synthesis apparatus, details for the crystal growth process are given in the Experimental Section. To identify the ambient-pressure structure of $Sm_4Ni_3O_{10-\delta}$, single-crystal X-ray diffraction (SXRD) measurements are performed on the as grown crystal with size around 0.097mm×0.095mm×0.035mm. Detailed crystallographic data for $Sm_4Ni_3O_{10-\delta}$ are summarized in Table 1. The SXRD data reveals an orthorhombic structure with space group *Pbca* (No. 61), showing a higher symmetry than the characteristic monoclinic structure with space group $P2_1/a$ (No. 14) observed in other $Ln_4Ni_3O_{10}$ trilayer nickelates. $Sm_4Ni_3O_{10-\delta}$ crystallizes in RP phase with n=3 (**Fig. 1a**), featuring three-layer $SmNiO_3$ perovskite blocks separated by SmO rock-salt layers along [001]. The presence of twinning was also identified in the single crystals, which is common in nickelates. The final single-crystal structure refinement was refined with a twin model. Although it consistently explains the essential diffraction features and reliably determines the fundamental lattice symmetry, it is important to acknowledge the limitations of this approach as an approximate representation of the real, more complex structure. More detailed descriptions of the precession images and the discussion about the twinning are provided in the supporting information. The Ni−O−Ni bond angle along the *c* axis is about 152.4(10)°, which is illustrated in the right panel of Fig. 1a. This significant deviation from 180° indicates pronounced distortion and tilting of the $NiO_6$ octahedra. In Table 1, it is noted that the largest difference peak/hole (3.140/-5.090) is close to the Sm atom, whose coordinate is (-0.0086(2), 0.9477(2), 0.06871(4)). And no disorder is found during the refinement modeled in the *Pbca* space group. The observed elevated residual electron density is primarily attributed to the effect of Fourier truncation errors, which is commonly encountered in structures containing heavy atoms.

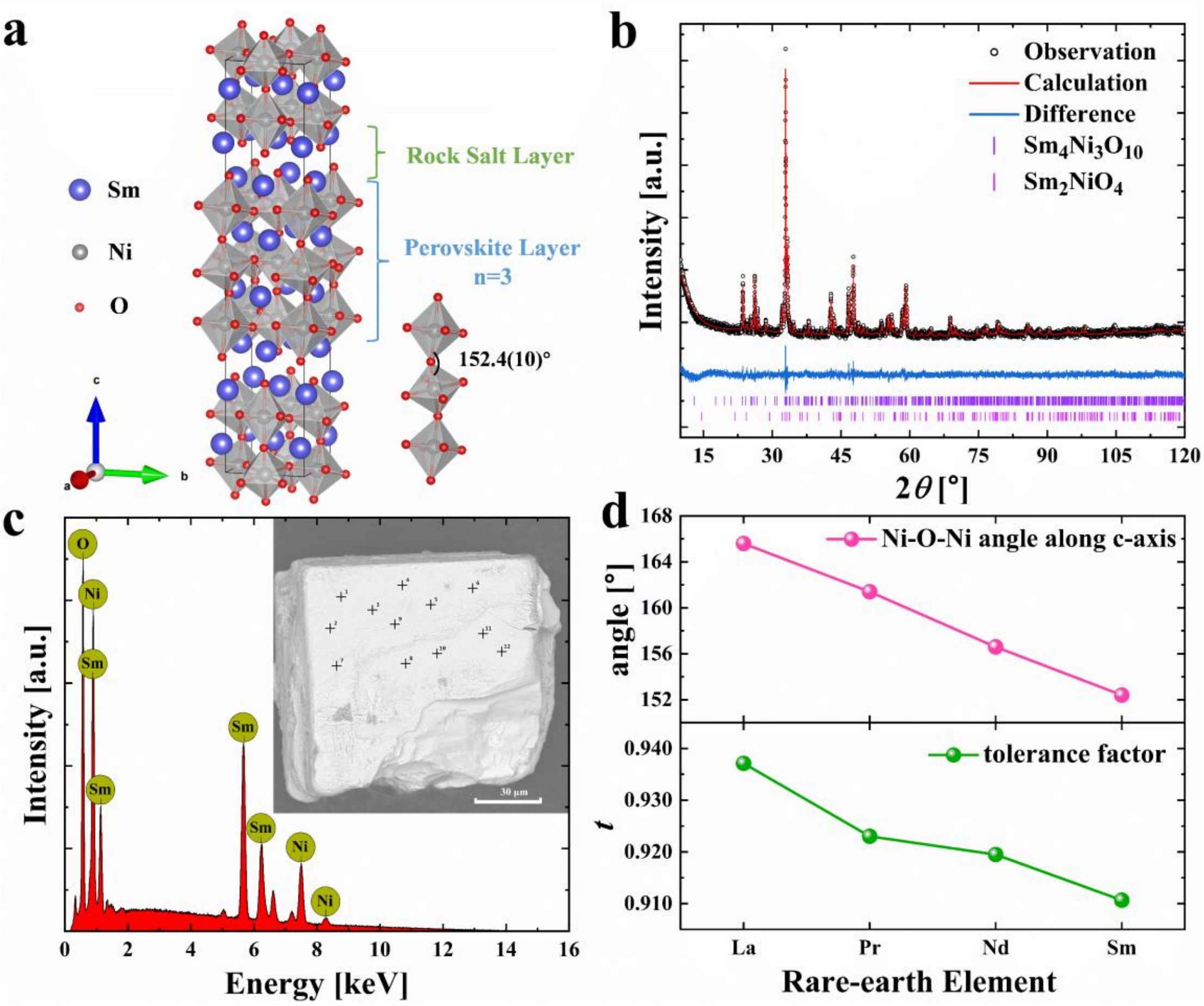

**Figure 1. Structural and compositional analysis of $Sm_4Ni_3O_{10-\delta}$ single crystals.** (a) Crystal structure of $Sm_4Ni_3O_{10-\delta}$ at ambient pressure. The Ni−O−Ni bond angle along the *c* axis indicates the distortion of the $NiO_6$ octahedra. (b) Rietveld refinements of PXRD pattern of $Sm_4Ni_3O_{10-\delta}$ at room temperature. (c) Energy dispersive spectrum (EDS) for the $Sm_4Ni_3O_{10-\delta}$ single crystal. The right inset illustrates the SEM image of the single crystal with a scale bar of 30 μm. (d) The Ni−O−Ni bond angle along the *c* axis and the tolerance factor *t* of $Ln_4Ni_3O_{10-\delta}$ (*Ln* = La, Pr, Nd, Sm) as a variation of rare-earth elements. Note that the *c*-axis Ni−O−Ni bond angles of $Ln_4Ni_3O_{10-\delta}$ (*Ln* = La, Pr, Nd) are collected from refs. [42, 45, 54], respectively.

To further verify the crystal structure of $Sm_4Ni_3O_{10-\delta}$, we performed powder X-ray diffraction (PXRD) by grinding the as grown single crystals into powder. Rietveld refinements were carried out using both the monoclinic $P2_1/a$ (Z=4) [35, 39, 45] and orthorhombic *Pbca* (Z=4) structural models. Figure 1b shows the refinement profile for the orthorhombic *Pbca* model, while the corresponding refinement for the monoclinic $P2_1/a$ (Z=4) model is provided in Fig. S1. And the detailed Rietveld refinement parameters are summarized in Table S1. Moreover, the Figure S2 displayed a direct comparison with the PXRD experimental data. Based on these refinements and structural analysis, we find that the structure of $Sm_4Ni_3O_{10-\delta}$ adopts the orthorhombic structure with space group *Pbca*. This assignment is supported by the inability of the monoclinic $P2_1/a$ (Z=4) model to accurately refine the peak intensities at 26.8°, 34.6° and 48.8°, whereas the orthorhombic *Pbca* (Z=4) model ($Ca_4Mn_3O_{10}$ type) provides excellent agreement with both position and intensity [50]. Additionally, two weak peaks observed at 24.5° and 32.3° (Fig. 1b) can be attributed to an impurity phase of $Sm_2NiO_4$ with a content of 9.827 wt.% from Rietveld refinement. This impurity phase may be formed because of the competition between the 214 ($Sm_2NiO_4$) and 4310 ($Sm_4Ni_3O_{10}$) Ruddlesden-Popper (RP) phases during crystal growth, which is a process extremely sensitive to oxygen partial pressure and temperature. Following this route, we intentionally change the amount of the external oxygen source $KClO_4$ during the growth, trying to lower down the impurity phases of $Sm_2NiO_4$ or $SmNiO_3$, and promote the formation of the main phase of $Sm_4Ni_3O_{10-\delta}$.

The inset in Fig. 1c displays a scanning electron microscopy (SEM) image of a representative $Sm_4Ni_3O_{10-\delta}$ single crystal with size of about 100 × 100 × 40 μm³. Prior to taking the image, the crystals were thoroughly washed with deionized water and anhydrous ethanol to remove residual KCl flux. The observed rectangular morphology and smooth surface indicate high crystal quality. Chemical composition analysis via energy-dispersive X-ray spectroscopy (EDS) yields a Sm: Ni atomic ratio of 4: 2.82 (Fig. 1c). This slight deviation from the nominal Sm: Ni ratio of 4: 3 in $Sm_4Ni_3O_{10-\delta}$ may arise from the EDS measurement limitations.

The Goldschmidt tolerance factor [51], $t = \frac{r_{Ln}+r_O}{\sqrt{2}(r_{Ni}+r_O)}$ (where $r_{Ln}$, $r_{Ni}$ and $r_O$ denote the effective ionic radii of the rare-earth, nickel, and oxygen ions, respectively [52,53]), is an empirical parameter widely used to assess structural stability and quantify octahedral distortion in perovskite-related oxides [47, 48]. Figure 1d illustrates the evolution trends of the Ni−O−Ni bond angle along the *c* axis and the tolerance factor *t* for $Ln_4Ni_3O_{10}$ (*Ln* = La, Pr, Nd, Sm). The tolerance factor decreases systematically with reducing rare-earth ionic radius. This intensified deviation from the ideal value ($t = 1$) significantly reduces structural stability. Consequently, synthesizing these compounds becomes progressively more challenging, requiring more precise oxygen pressures, narrower growth temperature windows, or elevated synthesis pressures. We find that HPHT methods combined with precise oxygen partial pressure control are essential for growing high-quality $Sm_4Ni_3O_{10-\delta}$ single crystals. Meanwhile, the Ni−O−Ni bond angle at ambient pressure along the *c* axis decreases consistently from a maximum of 165.6(6)° in $La_4Ni_3O_{10-\delta}$ to a minimum of 152.4(10)° in $Sm_4Ni_3O_{10-\delta}$, indicating increased distortion and tilting of the $NiO_6$ octahedra [42, 45, 54]. A more detailed discussion of the $NiO_6$ octahedra tilting is provided in the Supporting Information, as illustrated in Figure S4. These distortions prevent the formation of a straight ($\theta = 180°$) $c$-axis Ni−O−Ni configurations in $Sm_4Ni_3O_{10-\delta}$, creating intrinsic barriers to superconductivity.

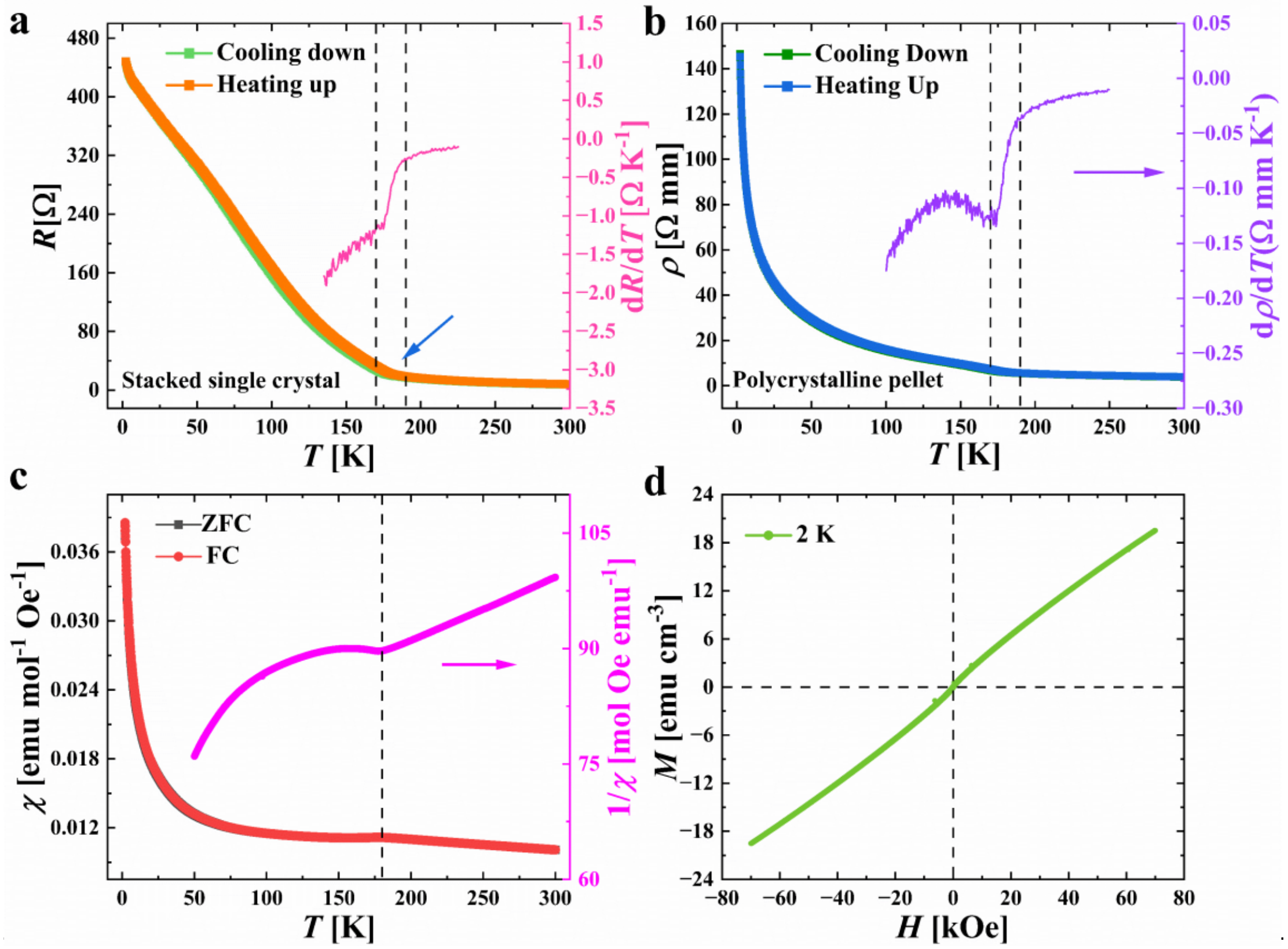

**Figure 2. Electrical transport and magnetization properties of $Sm_4Ni_3O_{10-\delta}$.** (a) The temperature-dependent resistance curve of the $Sm_4Ni_3O_{10-\delta}$ stacked single crystals. A sudden increase in resistance at approximately 180 K indicates the occurrence of a DW transition. (b) Temperature dependence of resistivity for the $Sm_4Ni_3O_{10-\delta}$ polycrystalline pellet sample at zero magnetic field and ambient pressure. Note that the polycrystalline pellets were treated at 500 °C in 8MPa $O_2$. (c)Temperature dependence of magnetic susceptibility measured at 20 kOe. The pink curve of $1/\chi$ versus $T$ shows an obvious transition consistent with transport measurements. (d) Magnetic hysteresis loop measured at 2 K, showing a paramagnetic or antiferromagnetic behavior.

To explore the electrical transport and magnetic properties of $Sm_4Ni_3O_{10-\delta}$, we conduct the temperature-dependent resistance measurements by using a physical property measurement system (Quantum Design, PPMS-9T) and the dc magnetization measurements by the vibrating sample magnetometer based on a superconducting interference device (Quantum Design, SQUID-VSM), the results are shown in **Fig 2**. Fig. 2a shows the electrical resistance $R(T)$ of the stacked $Sm_4Ni_3O_{10-\delta}$ single crystals. The stacked single crystals refer that different single crystals are naturally intergrown during the crystal growth process. A typical micrograph of "stacked single crystals" is provided in Figure S10. At ambient pressure, $Sm_4Ni_3O_{10-\delta}$ exhibits an increase in resistance as the temperature decreases from 300 K to 2 K. Especially, a notable anomalous kink is observed at approximately 180 K, below which the resistance shows an obvious upturn. Figure 2b presents the temperature dependent resistivity curve of the $Sm_4Ni_3O_{10-\delta}$ polycrystalline pellet, which is ground from single crystals and sintered at 500 °C under 8 MPa $O_2$ atmosphere. One can see that, the kink at ~180 K still exists, which is identified more distinctly with the derivative of the resistivity. Notably, this DW like transition was also observed in other trilayer nickelates, which can be attributed to the intertwined SDW and CDW [32-36]. In comparison, the DW like transition temperature of ~180 K in $Sm_4Ni_3O_{10-\delta}$ is higher than that in other trilayer nickelates which are 136 K (La), 156 K (Pr), and 162 K (Nd), respectively. Moreover, recent research has established a distinct correlation between the DW transition temperature and the Ni–O–Ni bond angle [55]. We have added the data point of $Sm_4Ni_3O_{10-\delta}$ into the phase diagram. As shown in Figure S6, the result of $Sm_4Ni_3O_{10}$ fits well with the reported trend. However, unlike other trilayer nickelates $Ln_4Ni_3O_{10-\delta}$ ($Ln$ = La, Pr, Nd), $Sm_4Ni_3O_{10-\delta}$ exhibits a weak insulating behavior in the range of 2-300 K, even after treatment under high oxygen pressure. And the DW transition looks more like a metal-to-insulator or weak insulator-to-insulator transition with an insulating ground state. In contrast, the La-Nd trilayer nickelates exhibit a metal-to-metal transition with a metallic ground state. This may be ascribed to the smaller Ni−O−Ni bond angle, which results in a significant $NiO_6$ octahedral tilt and distortion, as shown in Figure S4. The intensified octahedral tilt angle of $Sm_4Ni_3O_{10}$ will significantly reduce the overlap integral between Ni-3$d$ and O-2$p$ orbitals, resulting in higher resistivity and the insulating ground state.[17]. Additionally, oxygen deficiency can also lead to weak insulating behavior [10, 13, 14, 56].

The curves of temperature-dependent susceptibility of $Sm_4Ni_3O_{10-\delta}$ at 20 kOe detected in the zero-field-cooling (ZFC) and field-cooling (FC) process mode are shown in Fig. 2c. The magnetic susceptibility shows a Curie-Weiss like paramagnetic behavior in the temperature range from 180 K to 300 K. Moreover, it also exhibits an anomalous magnetic transition at ~180 K, which is more obvious in the form of $1/\chi$ versus $T$, corresponding to the DW transition observed in the electrical measurements. The magnetic susceptibility exhibits the Curie-Weiss like paramagnetic behavior above the DW transition temperature (180 K). A more detailed analysis of Curie-Weiss fit is provided in the supporting information Fig.S12. The magnetization hysteresis loop of $Sm_4Ni_3O_{10-\delta}$ at 2 K within $H = \pm 70$ kOe is shown in Fig. 2d. One can see that the $M(H)$ curve exhibits a paramagnetic or antiferromagnetic signal without a magnetic hysteresis loop, and the magnetization is unsaturated up to 70 kOe. Moreover, a weak nonlinearity near $H = 0$ Oe was observed. We thus performed additional low-field magnetization measurements at $H = 100$ Oe, which exhibited a spin-glass like behavior. In addition, the DW transition at 180 K may be accompanied by the emergence of antiferromagnetic order. A slight canting of the antiferromagnetically aligned spins will rise to a small net magnetic moment, manifesting as the observed low-field nonlinearity. Therefore, we ascribe the weak nonlinearity near $H$ = 0 Oe to spin-glass behavior or spin canting. More details are provided in Supporting Information.

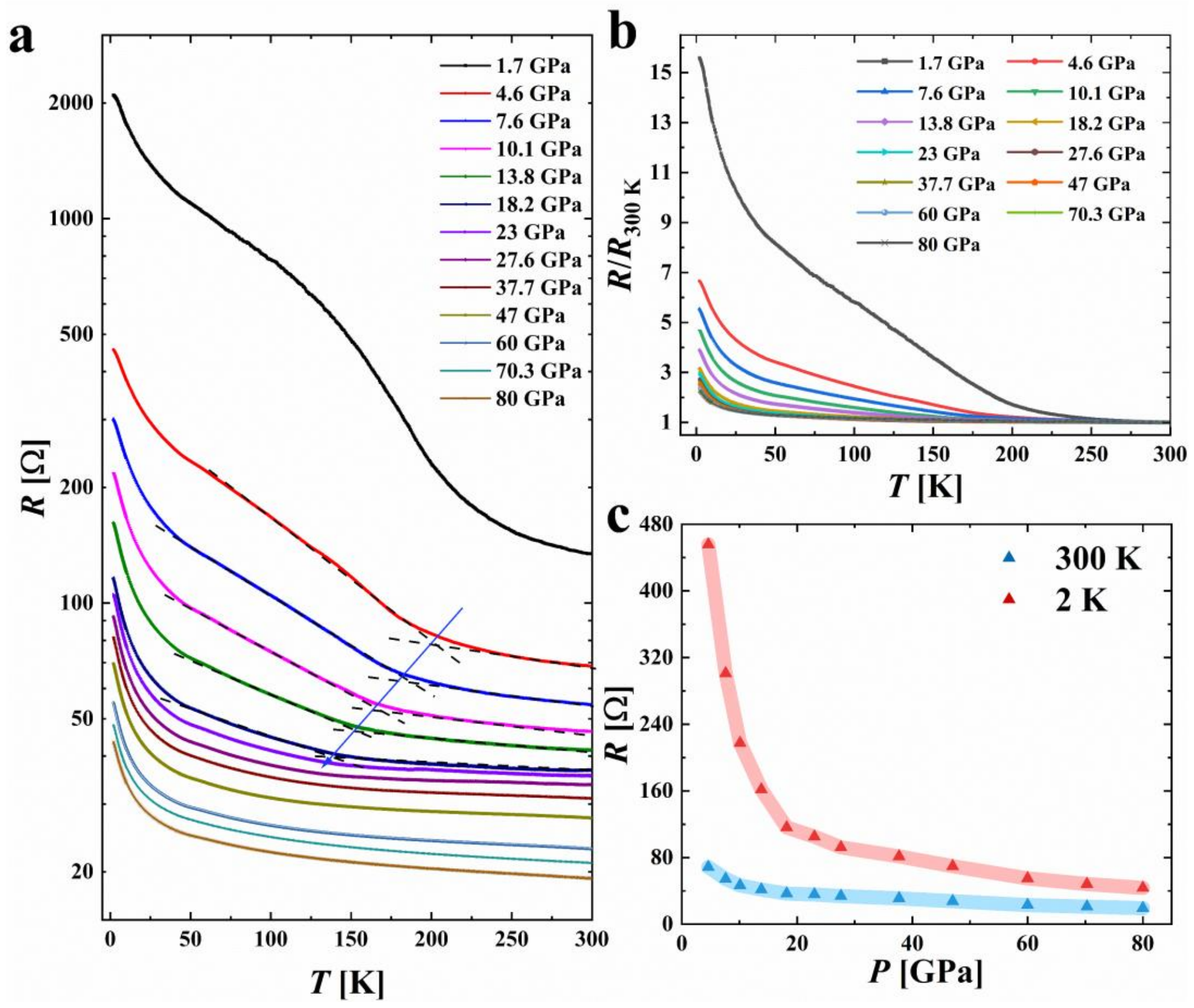

**Figure 3. High-pressure electrical resistance measurements on $Sm_4Ni_3O_{10-\delta}$ single crystals in a diamond-anvil-cell (DAC) apparatus.** (a) Temperature dependence of resistance of $Sm_4Ni_3O_{10-\delta}$ under various pressures from 1.7 GPa to 80 GPa. (b) Normalized $R/R_{300\,K}$ versus $T$ curves from 2 to 300 K. (c) Pressure dependent resistance ranging from 4.6 GPa to 80 GPa at 2 K and 300 K.

To explore the evolution of electrical transport properties and possible superconductivity in $Sm_4Ni_3O_{10-\delta}$ single crystals under pressure, the temperature-dependent resistance measurements were performed up to 80 GPa using diamond anvil cells (DAC). In **Fig. 3a**, the $R(T)$ curve at 1.7 GPa exhibits a weak insulating behavior throughout the entire temperature range with a DW transition. As the pressure gradually increases from 1.7 GPa to 18 GPa, the magnitude of $R(T)$ decreases monotonically over the entire temperature range, accompanied by an anomalous increase in resistance which may relate to the intertwined SDW and CDW. Furthermore, the evolution of the resistance anomaly is illustrated by an arrow and guidelines in Fig. 3a, which becomes almost invisible above 18.2 GPa, which is similar to that observed in $La_4Ni_3O_{10-\delta}$ and $Pr_4Ni_3O_{10-\delta}$ [42, 45]. Upon further compression, the $R(T)$ curves gradually transition from a weak insulating behavior to a semiconducting-like behavior and no superconductivity has been observed up to 80 GPa. As illustrated in Fig. 3b, based on the dependence of the resistance ratio $R/R_{300\ K}$ on temperature under pressure, it can be observed that as pressure increases, the upward trend of resistance is progressively suppressed, indicating that its insulating behavior is also suppressed by pressure. Figure 3c shows the pressure dependence of resistance at 2 K and 300 K over the pressure range from 4.6 GPa to 80 GPa. The resistance at both high and low temperatures decreases rapidly up to 20 GPa, as illustrated in Fig. 3b and 3c. This behavior may be related to atomic lattice compression or reduction of the $NiO_6$ octahedral distortion induced by pressure, which in turn modifies the electronic band structure, leading to an enhanced effective density of states at the Fermi energy.

However, superconductivity does not emerge with pressure up to 80 GPa in $Sm_4Ni_3O_{10-\delta}$ single crystals. The absence of superconductivity may be caused by the smaller Ni−O−Ni bond angle along the *c* axis, which significantly deviates further from 180°. In previous studies of pressurized bilayer and trilayer nickelates, the emergence of superconductivity is accompanied by a pressure-induced structural phase transition, namely the disappearance of $NiO_6$ octahedral distortion [42, 44, 45, 57-59]. Moreover, the pressure at which the structural phase transition from monoclinic to tetragonal occurs in $La_4Ni_3O_{10-\delta}$ is approximately 13-15 GPa [42], whereas that in $Pr_4Ni_3O_{10-\delta}$ exceeds over 35 GPa [45]. Following this line, because of the more severe structural distortion in $Pr_4Ni_3O_{10-\delta}$, the onset pressure for superconductivity is significantly higher than that in $La_4Ni_3O_{10-\delta}$. This indicates that as the Ni−O−Ni bond angle decreases from 180°, a higher pressure is required to induce the structural phase transition and achieve superconductivity. Therefore, considering that the Ni−O−Ni bond angle in $Sm_4Ni_3O_{10-\delta}$ is only 152.4(10)° at ambient pressure, which deviates substantially from 180°, a considerably higher pressure is expected for the structural phase transition or even the appearance of superconductivity. Therefore, synchrotron X-ray diffraction and electrical transport measurements under significantly higher pressures will be necessary to explore the possible high-temperature superconductivity in future studies. Besides, homogeneous pressure is also crucial to achieve superconductivity in nickelates, especially in $La_5Ni_3O_{11}$ single crystals whose superconductivity was observed only by helium as pressure transmitting medium [31].

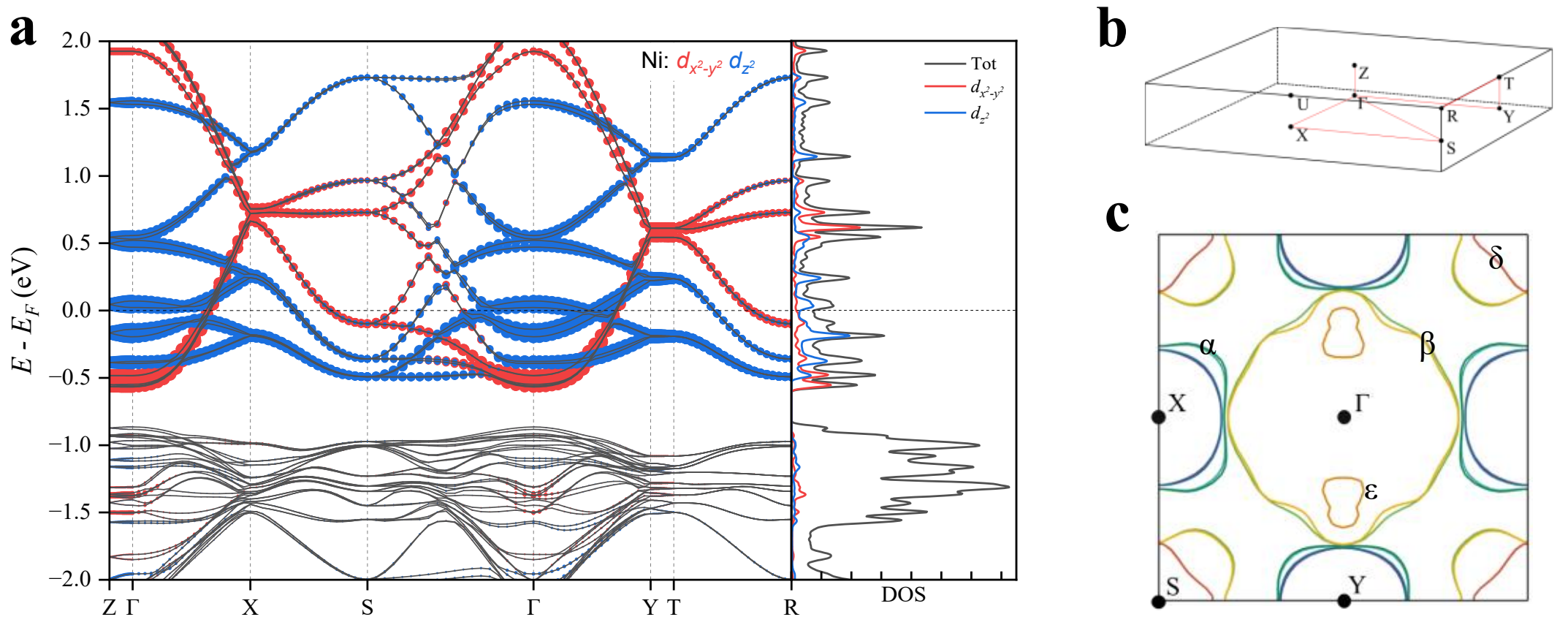

**Figure 4. DFT calculations for band structure and Fermi surface of $Sm_4Ni_3O_{10-δ}$ using the crystal structure determined at ambient pressure.** (a) The band structure of $Sm_4Ni_3O_{10-δ}$. The corresponding density of states (DOS) near the Fermi level are shown on the righthand side. The weight of $3d_{z^2}$ and $3d_{x^2-y^2}$ orbitals of Ni is represented by blue and red dots, respectively. Note that the dot size is proportional to the contribution of each orbital. (b) Schematic of the three-dimensional orthorhombic Brillouin zone. The red lines correspond to the paths of the electronic bands in (a). (c) The fermi surface of $Sm_4Ni_3O_{10-δ}$.

To investigate the electronic structure of $Sm_4Ni_3O_{10-δ}$, we perform density functional theory (DFT) calculation with the crystal structure determined at ambient pressure. **Figure 4a** shows the band structure of $Sm_4Ni_3O_{10-δ}$with the corresponding density of states (DOS) plotted on the right. The paths of the electronic bands are plotted with the red lines in three-dimensional orthorhombic Brillouin zone, as shown in Fig. 4b. There are multiple bands across the Fermi level, revealing a more complicated electronic band structure compared with that in $La_3Ni_2O_7$ [4]. The calculated results indicates that the Ni $e_g$ orbitals ($3d_{z^2}$ colored blue and $3d_{x^2-y^2}$ colored red) are separated from the $t_{2g}$ orbitals and exhibit a broad distribution across the Fermi level, driven by crystal-field splitting in the $NiO_6$ octahedra [4, 59]. Regarding the interlayer $\sigma$-bond formed by the hybridization of Ni $3d_{z^2}$ and O $2p_z$ orbitals, there are bonding, nonbonding, and antibonding bands below and above the Fermi level, respectively. The top of the bonding band is located near and even a little above the Fermi level, while the bottom of the antibonding band is placed completely above the Fermi level. Accordingly, it is illustrated that the nonbonding band is across the Fermi level. In previous studies, all three $3d_{z^2}$ bands are expected to contribute to superconductivity in $La_4Ni_3O_{10-δ}$ [40]. Given the similar relationship between the Fermi level and the band edges in $La_4Ni_3O_{10-δ}$, modifications such as applying pressure or carrier doping may facilitate the emergence of superconductivity in $Sm_4Ni_3O_{10-δ}$.

Figure 4c displays the calculated two-dimensional Fermi surface of $Sm_4Ni_3O_{10-\delta}$, which dominantly consists of $3d_{x^2-y^2}$ and $3d_{z^2}$ orbital components. The $3d_{x^2-y^2}$ band crosses the Fermi level and forms an electron pocket (*β* band) which is centered at the *Γ* point in the Fermi surface (Fig. 4c). The bonding and nonbonding bands intertwine around the Fermi level near the *Γ* point, leading to an additional small pocket (*ε* band) along the *Γ*-*Y* path as depicted in Fig. 4c. This discrepancy from that in $La_4Ni_3O_{10-\delta}$ may be ascribed to the intensified $NiO_6$ octahedral distortion in orthorhombic structure of $Sm_4Ni_3O_{10-\delta}$, which weakens the interlayer *σ*-bond coupling between the Ni $3d_{z^2}$ and O $2p_z$ orbitals. Additionally, in considering the strong electronic correlation in trilayer nickelates, we performed the DFT+U calculations as shown in Figure S7. It is found that the Ni- $3d_{z^2}$ and Ni- $3d_{x^2-y^2}$ orbitals cross the Fermi level and exhibit a broader distribution. Therefore, distinct broadenings of both the *α* band (at the Brillouin zone side) and the *β* band (at the Brillouin zone corner) are observed. A more detailed discussion is provided in the Supporting Information.

**Table 1. Crystallographic data for $Sm_4Ni_3O_{10-\delta}$ obtained from the room-temperature SXRD.**

| | |
|---|---|
| Empirical formula | $Sm_4Ni_3O_{10-\delta}$ |
| Crystal system | orthorhombic |
| Space group | *P b c a* |
| Temperature [K] | 304.62(10) |
| Formula weight | 937.53 |
| a [Å] | 5.3844(4) |
| b [Å] | 5.4062(5) |
| c [Å] | 27.287(2) |
| α [°] | 90 |
| β [°] | 90 |
| γ [°] | 90 |
| Volume [$Å^3$] | 794.30(11) |
| Density (calculated) [g $cm^{-3}$] | 7.840 |
| Z | 4 |
| Radiation type | Mo-$K_\alpha$ (λ = 0.71073 Å) |
| Crystal size [$mm^3$] | 0.097×0.095×0.035 |
| Absorption coefficient [$mm^{-1}$] | 36.074 |
| Data collection diffractometer | XtaLAB Synergy R, DW system, HyPix |
| Absorption correction | Multi-scan |
| Reflections collected | 2834 |
| Independent reflections | 873 ($R_{int}$ = 0.0222) |
| θ range for data collection [°] | 2.239 –26.366 |
| *F* (000) | 1648 |
| Index ranges | -6≤h≤6, -6≤k≤6, -34≤l≤30 |
| Data, restraints, parameters | 873, 6, 80 |
| Goodness of fit on F2 | 1.303 |
| Final R indexes (I > 2σ) | $R_1$=0.0510, $wR_2$=0.1236 |
| Final R indexes (all data) | $R_1$=0.0552, $wR_2$=0.1258 |
| Largest diff. peak/hole [e Å$^{-3}$] | 3.140/-5.090 |

| **Site label** | **x** | **y** | **z** | **$U_{(eq)}$** | **$U_{11}$** | **$U_{22}$** | **$U_{33}$** | **$U_{23}$** | **$U_{13}$** | **$U_{12}$** |
|---|---|---|---|---|---|---|---|---|---|---|
| Sm1 | -0.0105(2) | 1.0246(2) | 0.20033(4) | 16.1(3) | 13.2(6) | 19.6(7) | 15.6(6) | -0.1(4) | 1.3(4) | 1.4(5) |
| Sm2 | -0.0086(2) | 0.9477(2) | 0.06871(4) | 15.3(3) | 11.6(6) | 19.7(7) | 14.6(5) | -0.8(5) | 0.1(4) | 0.8(5) |
| Ni1 | 0.0030(5) | 0.4914(6) | 0.13969(10) | 13.2(6) | 7.5(14) | 17.0(17) | 15.1(12) | -1.6(10) | -0.2(10) | 0.0000(12) |
| Ni2 | 0 | 0.5 | 0 | 14.1(8) | 10(2) | 17(2) | 15.7(17) | -2.0(15) | -0.4(15) | -1.2(17) |
| O1 | -0.066(3) | 0.477(4) | 0.2180(6) | 22(4) | 23(9) | 23(10) | 18(8) | 4(7) | -1(7) | 7(8) |
| O2 | -0.2170(3) | 1.213(3) | 0.1297(5) | 17(3) | 14(8) | 16(8) | 20(7) | 4 (7) | 4(7) | -5 (7) |
| O3 | -0.295(3) | 0.706(3) | 0.0130(5) | 17(3) | 13(8) | 21(9) | 16(7) | -1(7) | 1(7) | -1(7) |
| O4 | 0.414(3) | 1.016(4) | 0.0679(5) | 21(4) | 18(8) | 32(11) | 13(7) | -7(8) | 6(6) | -5(8) |
| O5 | 0.220(3) | 0.771(3) | 0.1514(5) | 19(3) | 19(4) | 18(4) | 19(4) | -1(3) | -1(3) | -2(3) |

The unit of the equivalent isotropic ($U_{eq}$) and anisotropic ($U_{ij}$) displacement parameters is 0.001 Å$^2$

## 3. Conclusions

In summary, we have successfully synthesized the new trilayer RP nickelate $Sm_4Ni_3O_{10-\delta}$ single crystals using an apparatus with HPHT techniques. Structural and compositional analysis confirms the trilayer orthorhombic structure (*Pbca*), and the significant deviation of the apical oxygen bond angle (Ni−O−Ni) from 180°. Magnetization and electric transport measurements at ambient pressure consistently reveal a density wave transition at approximately 180 K. However, no superconductivity is observed in high-pressure transport measurements up to 80 GPa. The stronger distortion of $NiO_6$ octahedra may account for the absence of superconductivity. DFT calculations indicate the separation of $e_g$ orbitals and $t_{2g}$ orbitals in electronic band structure. Compared with the Fermi surface of $La_4Ni_3O_{10-\delta}$, a slightly difference indicates the distortion affects the electronic band structure. For the expansion of the trilayer RP nickelate family, the HPHT flux growth method has been demonstrated as an innovative approach to overcome the limitations of ambient-pressure RP nickelate synthesis. Our findings on $Sm_4Ni_3O_{10-\delta}$ suggest that the distortion mediated by tolerance factor serve as a key design parameter for tailoring electronic states and superconductivity, which offers a broader perspective for exploring superconductivity in more nickelates with RP phases.

## 4. Experimental Section

### 4.1 Single crystal growth and characterization

The precursor of $Sm_4Ni_3O_{10-\delta}$ was synthesized via a modified sol-gel method. Stoichiometric amounts of $Sm_2O_3$ (99.9%, Alfa Aesar, dried at 1000°C prior to use) and $Ni(OH)_2$ (99%, Picasso) were weighed in the glove box and dissolved in nitric acid. And the mixture was heated to 90 °C to facilitate dissolution. Then an equivalent molar proportion of citric acid with respect to cations was added to prevent precipitating prematurely. Some ethylene glycol was also added to achieve a more homogeneous state. After several hours heating, the green gel product was formed and transferred into muffle furnace, heating at 500 °C for 10 h. Then the dark grey product was ground to powder and sintered in an alumina crucible at 950 °C in air for 10 h. Now the precursor was highly homogeneous, without organic compounds and nitrogen oxides.

The $Sm_4Ni_3O_{10-\delta}$ single crystals were grown via a flux method with KCl as a flux, using the HPHT apparatus. The $Sm_4Ni_3O_{10-\delta}$ precursor, KCl (99.8%, Aladdin) and $KClO_4$ (99%, Alfa Aesar) were weighed in a molar ratio of 1: 2.5: 0.33. In HPHT synthesis, $KClO_4$ is a widely used oxygen source. Its decomposition product (KCl) is chemically identical to the flux, thereby avoiding the additional impurity phases. To systematically optimize the oxygen content, we carefully adjusted the amount of $KClO_4$ and analysized the resulting phase formation. As summarized in the table S2, the molar ratio of $KClO_4$ effectively modulates the final product. A molar ratio of 0.33:1 yields the trilayer $Sm_4Ni_3O_{10}$ phase with the highest purity and optimal crystal quality. This systematic optimization not only ensures high reproducibility but also validates the controlled synthesis of the target compound. The mixture was completely ground for 20 min and pressed into pellets to seal in a platinum capsule. All the above procedures were carried out inside a glove box with $O_2$ and $H_2O$ content less than 1 ppm. Then the capsule was heated at 1400 °C under 3.25 GPa for 4 h using a piston-cylinder-type high-pressure apparatus (LP 1000–540/50, Max Voggenreiter). After that, the temperature was quenched to room temperature and then the high pressure was subsequently released.

*Crystal Structural Determination and Sample Characterization:* The crystal structure of $Sm_4Ni_3O_{10-\delta}$ was identified by single-crystal XRD (Rigaku, XtaLAB Synergy-DW) with Mo-$K_\alpha$ radiation ($\lambda$ = 0.71073 Å) (Centre for Shared Scientific Research Facilities, Nanjing University). The data collected at 300 K was reduced and finalized using CrysAlis$^{Pro}$ software and the structure were solved and refined using Olex2 with ShelXT and ShelXL packages [60-63]. The powder XRD was conducted with a Bruker D8 Advance diffractometer with Cu-$K_\alpha$ radiation ($\lambda$ = 1.54184 Å) in the 2θ range from 10° to 120°. TOPAS 4.2 was used in Rietveld refinements [64], where the single-crystal structure model served as the starting model. Single crystal micrographs were obtained from a scanning electron microscope (SEM, Phenom ProX), and the chemical composition analysis was conducted via the energy dispersive X-ray spectrometer (EDS) equipped on the SEM. Note that more details of the crystal structure for $Sm_4Ni_3O_{10-\delta}$ can be obtained from the joint CCDC/FIZ Karlsruhe online deposition service by quoting the deposition number CSD 2479415.

### 4.2 Magnetization and resistivity measurements

The dc magnetization measurements were performed with the superconducting quantum interference device (SQUID-VSM-7 T, Quantum Design). Temperature-dependent resistance measurements for stacked single crystals and polycrystalline pellets were conducted on a physical property measurement system (PPMS-9T, Quantum Design). The polycrystalline pellets were obtained by grinding microcrystals and pressing the resulting powder. Additionally, an annealing treatment at 500 °C under an 8 MPa $O_2$ atmosphere was performed to enhance interparticle connectivity.

High-pressure transport measurements: Diamond anvil cell (DACPPMS-ET225, Shanghai Anvil source Material Technology Co., Ltd) with a 200 μm culet was used to generate pressures up to 80 GPa. And the four-probe van der Pauw method was applied for the high-pressure resistance measurements with KBr serving as the pressure-transmitting medium. Pressures were calibrated by the ruby fluorescence method at room temperature.

### 4.3 DFT calculations

The DFT calculations were performed using the plane-wave DFT within a projector-augmented wave scheme as implemented in the Vienna Ab initio Simulation Package (VASP) [65-68]. We used the generalized gradient approximation in the Perdew-Burke-Ernzerhof form in the self-consistent calculation of the electronic structure [69]. A $9 \times 9 \times 2$ k-point grid with a cutoff energy of 580 eV was set, and the lattice constants were taken from the structure identified by single-crystal XRD.

### Author Contributions

The growth of $Sm_4Ni_3O_{10-\delta}$ single crystals, SEM and EDS analyses were performed by Y.Z., T.-Y. L., and X.Z. The SXRD and PXRD data were collected by Y.Z. and X.Z. The resistivity and magnetization measurements at ambient pressure were done by Y.Z. and T.-Y. L. The high-pressure electrical resistance measurements were conducted by Y.-J. Z and Q.L. The DFT calculations was finished by S.F. The paper was written by X.Z., Y.Z., Q. L. and H.-H.W. All authors joined the analysis and agreed to publish the data. H.-H.W. has coordinated the whole work.

### Acknowledgements

This work was supported by the National Key Research and Development Program of China (No. 2022YFA1403201), National Natural Science Foundation of China (Nos. 1243000380, 123B2055, 52472276, 12061131001, 11927809, 12494591), Fundamental Research Funds for the Central Universities (Grant No. 2024300350). We thank the Centre for Shared Scientific Research Facilities of Nanjing University for support, and Professor Guiling Shi for her assistance with the single-crystal XRD measurements. We thank Dr. Jin Wu from Rigaku Shanghai Corporation (RSHC) for assistance with the single-crystal structure determination.

**Declaration of competing interest**

The authors declare that they have no known competing financial interests or personal relationships that could have appeared to influence the work reported in this paper.

**Data Availability Statement**

Data will be made available on request.

**References**

[1] J.G. Bednorz, K.A. Müller, Possible high $T_c$ superconductivity in the Ba−La−Cu−O system, Z. Phys. B Condens. Matter 64 (1986) 189–193. https://doi.org/10.1007/BF01303701.

[2] Y. Kamihara, T. Watanabe, M. Hirano, H. Hosono, Iron-based layered superconductor La[$O_{1-x}F_x$]FeAs (x = 0.05−0.12) with $T_c$ = 26 K, J. Am. Chem. Soc. 130 (2008) 3296–3297. https://doi.org/10.1021/ja800073m.

[3] D. Li, K. Lee, B.Y. Wang, M. Osada, S. Crossley, H.R. Lee, Y. Cui, Y. Hikita, H.Y. Hwang, Superconductivity in an infinite-layer nickelate, Nature 572 (2019) 624–627. https://doi.org/10.1038/s41586-019-1496-5.

[4] H. Sun, M. Huo, X. Hu, J. Li, Z. Liu, Y. Han, L. Tang, Z. Mao, P. Yang, B. Wang, J. Cheng, D.-X. Yao, G.-M. Zhang, M. Wang, Signatures of superconductivity near 80 K in a nickelate under high pressure, Nature 621 (2023) 493–498. https://doi.org/10.1038/s41586-023-06408-7.

[5] V.I. Anisimov, D. Bukhvalov, T.M. Rice, Electronic structure of possible nickelate analogs to the cuprates, Phys. Rev. B 59 (1999) 7901–7906. https://doi.org/10.1103/PhysRevB.59.7901.

[6] K. Lee, B.Y. Wang, M. Osada, B.H. Goodge, T.C. Wang, Y. Lee, S. Harvey, W.J. Kim, Y. Yu, C. Murthy, S. Raghu, L.F. Kourkoutis, H.Y. Hwang, Linear-in-temperature resistivity for optimally superconducting (Nd,Sr)$NiO_2$, Nature 619 (2023) 288–292. https://doi.org/10.1038/s41586-023-06129-x.

[7] Q. Gu, H.-H. Wen, Superconductivity in nickel-based 112 systems, The Innovation 3 (2022) 100202. https://doi.org/10.1016/j.xinn.2021.100202.

[8] G.A. Pan, D. Ferenc Segedin, H. LaBollita, Q. Song, E.M. Nica, B.H. Goodge, A.T. Pierce, S. Doyle, S. Novakov, D. Córdova Carrizales, A.T. N'Diaye, P. Shafer, H. Paik, J.T. Heron, J.A. Mason, A. Yacoby, L.F. Kourkoutis, O. Erten, C.M. Brooks, A.S. Botana, J.A. Mundy, Superconductivity in a quintuple-layer square-planar nickelate, Nat. Mater. 21 (2022) 160–164. https://doi.org/10.1038/s41563-021-01142-9.

[9] S.L.E. Chow, Z. Luo, A. Ariando, Bulk superconductivity near 40 K in hole-doped $SmNiO_2$ at ambient pressure, Nature 642 (2025) 58–63. https://doi.org/10.1038/s41586-025-08893-4.

[10] Z. Dong, M. Huo, J. Li, J. Li, P. Li, H. Sun, L. Gu, Y. Lu, M. Wang, Y. Wang, Z. Chen, Visualization of oxygen vacancies and self-doped ligand holes in $La_3Ni_2O_{7-\delta}$, Nature 630 (2024) 847–852. https://doi.org/10.1038/s41586-024-07482-1.

[11] G. Wang, N.N. Wang, X.L. Shen, J. Hou, L. Ma, L.F. Shi, Z.A. Ren, Y.D. Gu, H.M. Ma, P.T. Yang, Z.Y. Liu, H.Z. Guo, J.P. Sun, G.M. Zhang, S. Calder, J.-Q. Yan, B.S. Wang, Y. Uwatoko, J.-G. Cheng, Pressure-induced superconductivity in polycrystalline $La_3Ni_2O_{7-\delta}$, Phys. Rev. X 14 (2024) 011040. https://doi.org/10.1103/PhysRevX.14.011040.

[12] Y. Zhang, D. Su, Y. Huang, Z. Shan, H. Sun, M. Huo, K. Ye, J. Zhang, Z. Yang, Y. Xu, Y. Su, R. Li, M. Smidman, M. Wang, L. Jiao, H. Yuan, High-temperature superconductivity with zero resistance and strange-metal behaviour in $La_3Ni_2O_{7-\delta}$, Nat. Phys. 20 (2024) 1269–1273. https://doi.org/10.1038/s41567-024-02515-y.

[13] M. Wang, H.-H. Wen, T. Wu, D.-X. Yao, T. Xiang, Normal and superconducting properties of $La_3Ni_2O_7$, Chin. Phys. Lett. 41 (2024) 077402. https://doi.org/10.1088/0256-307X/41/7/077402.

[14] Z. Zhang, M. Greenblatt, J.B. Goodenough, Synthesis, structure, and properties of the layered perovskite $La_3Ni_2O_{7-\delta}$, J. Solid State Chem. 108 (1994) 402–409. https://doi.org/10.1006/jssc.1994.1059.

[15] Z. Pan, C. Lu, F. Yang, C. Wu, Effect of rare-earth element substitution in superconducting $R_3Ni_2O_7$ under Pressure, Chin. Phys. Lett. 41 (2024) 087401. https://doi.org/10.1088/0256-307X/41/8/087401.

[16] B. Geisler, J.J. Hamlin, G.R. Stewart, R.G. Hennig, P.J. Hirschfeld, Structural transitions, octahedral rotations, and electronic properties of $A_3Ni_2O_7$ rare-earth nickelates under high pressure, Npj Quantum Mater. 9 (2024) 38. https://doi.org/10.1038/s41535-024-00648-0.

[17] N. Wang, G. Wang, X. Shen, J. Hou, J. Luo, X. Ma, H. Yang, L. Shi, J. Dou, J. Feng, J. Yang, Y. Shi, Z. Ren, H. Ma, P. Yang, Z. Liu, Y. Liu, H. Zhang, X. Dong, Y. Wang, K. Jiang, J. Hu, S. Nagasaki, K. Kitagawa, S. Calder, J. Yan, J. Sun, B. Wang, R. Zhou, Y. Uwatoko, J. Cheng, Bulk high-temperature superconductivity in pressurized tetragonal $La_2PrNi_2O_7$, Nature 634 (2024) 579–584. https://doi.org/10.1038/s41586-024-07996-8.

[18] F. Li, Z. Xing, D. Peng, J. Dou, N. Guo, L. Ma, Y. Zhang, L. Wang, J. Luo, J. Yang, J. Zhang, T. Chang, Y.-S. Chen, W. Cai, J. Cheng, Y. Wang, Z. Zeng, Q. Zheng, R. Zhou, Q. Zeng, X. Tao, J. Zhang, Ambient pressure growth of bilayer nickelate single crystals with superconductivity over 90 K under high pressure, Preprint in arxiv (2025). https://doi.org/10.48550/arXiv.2501.14584.

[19] X. Chen, J. Zhang, A.S. Thind, S. Sharma, H. LaBollita, G. Peterson, H. Zheng, D.P. Phelan, A.S. Botana, R.F. Klie, J.F. Mitchell, Polymorphism in the Ruddlesden–Popper nickelate $La_3Ni_2O_7$: Discovery of a hidden phase with distinctive layer stacking, J. Am. Chem. Soc. 146 (2024) 3640–3645. https://doi.org/10.1021/jacs.3c14052.

[20] P. Puphal, P. Reiss, N. Enderlein, Y.-M. Wu, G. Khaliullin, V. Sundaramurthy, T. Priessnitz, M. Knauft, A. Suthar, L. Richter, M. Isobe, P.A. Van Aken, H. Takagi, B. Keimer, Y.E. Suyolcu, B. Wehinger, P. Hansmann, M. Hepting, Unconventional crystal

structure of the high-pressure superconductor $La_3Ni_2O_7$, Phys. Rev. Lett. 133 (2024) 146002. https://doi.org/10.1103/PhysRevLett.133.146002.

[21] H. Wang, L. Chen, A. Rutherford, H. Zhou, W. Xie, Long-range structural order in a hidden phase of Ruddlesden–Popper bilayer nickelate $La_3Ni_2O_7$, Inorg. Chem. 63 (2024) 5020–5026. https://doi.org/10.1021/acs.inorgchem.3c04474.

[22] E.K. Ko, Y. Yu, Y. Liu, L. Bhatt, J. Li, V. Thampy, C.-T. Kuo, B.Y. Wang, Y. Lee, K. Lee, J.-S. Lee, B.H. Goodge, D.A. Muller, H.Y. Hwang, Signatures of ambient pressure superconductivity in thin film $La_3Ni_2O_7$, Nature 638 (2025) 935–940. https://doi.org/10.1038/s41586-024-08525-3.

[23] G. Zhou, W. Lv, H. Wang, Z. Nie, Y. Chen, Y. Li, H. Huang, W.-Q. Chen, Y.-J. Sun, Q.-K. Xue, Z. Chen, Ambient-pressure superconductivity onset above 40 K in $(La,Pr)_3Ni_2O_7$ films, Nature 640 (2025) 641–646. https://doi.org/10.1038/s41586-025-08755-z.

[24] Y. Liu, E.K. Ko, Y. Tarn, L. Bhatt, J. Li, V. Thampy, B.H. Goodge, D.A. Muller, S. Raghu, Y. Yu, H.Y. Hwang, Superconductivity and normal-state transport in compressively strained $La_2PrNi_2O_7$ thin films, Nat. Mater. 24 (2025) 1221–1227. https://doi.org/10.1038/s41563-025-02258-y.

[25] B. Hao, M. Wang, W. Sun, Y. Yang, Z. Mao, S. Yan, H. Sun, H. Zhang, L. Han, Z. Gu, J. Zhou, D. Ji, Y. Nie, Superconductivity in Sr-doped $La_3Ni_2O_7$ thin films, Nat. Mater. (2025). https://doi.org/10.1038/s41563-025-02327-2.

[26] P. Li, G. Zhou, W. Lv, Y. Li, C. Yue, H. Huang, L. Xu, J. Shen, Y. Miao, W. Song, Z. Nie, Y. Chen, H. Wang, W. Chen, Y. Huang, Z.-H. Chen, T. Qian, J. Lin, J. He, Y.-J. Sun, Z. Chen, Q.-K. Xue, Angle-resolved photoemission spectroscopy of superconducting $(La,Pr)_3Ni_2O_7/SrLaAlO_4$ heterostructures, Natl. Sci. Rev. (2025) nwaf205. https://doi.org/10.1093/nsr/nwaf205.

[27] J. Shen, G. Zhou, Y. Miao, P. Li, Z. Ou, Y. Chen, Z. Wang, R. Luan, H. Sun, Z. Feng, X. Yong, Y. Li, L. Xu, W. Lv, Z. Nie, H. Wang, H. Huang, Y.-J. Sun, Q.-K. Xue, J. He, Z. Chen, Nodeless superconducting gap and electron-boson coupling in $(La,Pr,Sm)_3Ni_2O_7$ films, Preprint in arXiv (2025). https://doi.org/10.48550/arXiv.2502.17831.

[28] B.Y. Wang, Y. Zhong, S. Abadi, Y. Liu, Y. Yu, X. Zhang, Y.-M. Wu, R. Wang, J. Li, Y. Tarn, E.K. Ko, V. Thampy, M. Hashimoto, D. Lu, Y.S. Lee, T.P. Devereaux, C. Jia, H.Y. Hwang, Z.-X. Shen, Electronic structure of compressively strained thin film $La_2PrNi_2O_7$, Preprint in arXiv (2025). https://doi.org/10.48550/arXiv.2504.16372.

[29] S. Fan, M. Ou, M. Scholten, Q. Li, Z. Shang, Y. Wang, J. Xu, H. Yang, I.M. Eremin, H.-H. Wen, Superconducting gap structure and bosonic mode in $La_2PrNi_2O_7$ thin films at ambient pressure, Preprint in arXiv (2025). https://doi.org/10.48550/arXiv.2506.01788.

[30] F. Li, N. Guo, Q. Zheng, Y. Shen, S. Wang, Q. Cui, C. Liu, S. Wang, X. Tao, G.-M. Zhang, J. Zhang, Design and synthesis of three-dimensional hybrid Ruddlesden-Popper nickelate single crystals, Phys. Rev. Mater. 8 (2024) 053401. https://doi.org/10.1103/PhysRevMaterials.8.053401.

[31] M. Shi, D. Peng, K. Fan, Z. Xing, S. Yang, Y. Wang, H. Li, R. Wu, M. Du, B. Ge, Z. Zeng, Q. Zeng, J. Ying, T. Wu, X. Chen, Pressure induced superconductivity in hybrid Ruddlesden–Popper $La_5Ni_3O_{11}$ single crystals, Nat. Phys. (2025). https://doi.org/10.1038/s41567-025-03023-3.

[32] J. Zhang, D. Phelan, A.S. Botana, Y.-S. Chen, H. Zheng, M. Krogstad, S.G. Wang, Y. Qiu, J.A. Rodriguez-Rivera, R. Osborn, S. Rosenkranz, M.R. Norman, J.F. Mitchell, Intertwined density waves in a metallic nickelate, Nat. Commun. 11 (2020) 6003. https://doi.org/10.1038/s41467-020-19836-0.

[33] A.M. Samarakoon, J. Strempfer, J. Zhang, F. Ye, Y. Qiu, J.-W. Kim, H. Zheng, S. Rosenkranz, M.R. Norman, J.F. Mitchell, D. Phelan, Bootstrapped dimensional crossover of a spin density wave, Phys. Rev. X 13 (2023) 041018. https://doi.org/10.1103/PhysRevX.13.041018.

[34] S. Huangfu, G.D. Jakub, X. Zhang, O. Blacque, P. Puphal, E. Pomjakushina, F.O. Von Rohr, A. Schilling, Anisotropic character of the metal-to-metal transition in $Pr_4Ni_3O_{10}$, Phys. Rev. B 101 (2020) 104104. https://doi.org/10.1103/PhysRevB.101.104104.

[35] B.-Z. Li, C. Wang, P.T. Yang, J.P. Sun, Y.-B. Liu, J. Wu, Z. Ren, J.-G. Cheng, G.-M. Zhang, G.-H. Cao, Metal-to-metal transition and heavy-electron state in $Nd_4Ni_3O_{10-\delta}$, Phys. Rev. B 101 (2020) 195142. https://doi.org/10.1103/PhysRevB.101.195142.

[36] Q. Li, C. He, X. Zhu, J. Si, X. Fan, H.-H. Wen, Contrasting physical properties of the trilayer nickelates $Nd_4Ni_3O_{10}$ and $Nd_4Ni_3O_8$, Sci. China Phys. Mech. Astron. 64 (2021) 227411. https://doi.org/10.1007/s11433-020-1613-3.

[37] G. Wang, N. Wang, T. Lu, S. Calder, J. Yan, L. Shi, J. Hou, L. Ma, L. Zhang, J. Sun, B. Wang, S. Meng, M. Liu, J. Cheng, Chemical versus physical pressure effects on the structure transition of bilayer nickelates, Npj Quantum Mater. 10 (2025) 1. https://doi.org/10.1038/s41535-024-00721-8.

[38] J. Zhang, H. Zheng, Y.-S. Chen, Y. Ren, M. Yonemura, A. Huq, J.F. Mitchell, High oxygen pressure floating zone growth and crystal structure of the metallic nickelates $R_4Ni_3O_{10}$ (R=La, Pr), Phys. Rev. Mater. 4 (2020) 083402. https://doi.org/10.1103/PhysRevMaterials.4.083402.

[39] M. Zhang, C. Pei, Q. Wang, Y. Zhao, C. Li, W. Cao, S. Zhu, J. Wu, Y. Qi, Effects of pressure and doping on Ruddlesden-Popper phases $La_{n+1}Ni_nO_{3n+1}$, J. Mater. Sci. Technol. 185 (2024) 147–154. https://doi.org/10.1016/j.jmst.2023.11.011.

[40] H. Sakakibara, M. Ochi, H. Nagata, Y. Ueki, H. Sakurai, R. Matsumoto, K. Terashima, K. Hirose, H. Ohta, M. Kato, Y. Takano, K. Kuroki, Theoretical analysis on the possibility of superconductivity in the trilayer Ruddlesden-Popper nickelate $La_4Ni_3O_{10}$ under pressure and its experimental examination: Comparison with $La_3Ni_2O_7$, Phys. Rev. B 109 (2024) 144511. https://doi.org/10.1103/PhysRevB.109.144511.

[41] Q. Li, Y.-J. Zhang, Z.-N. Xiang, Y. Zhang, X. Zhu, H.-H. Wen, Signature of superconductivity in pressurized $La_4Ni_3O_{10}$, Chin. Phys. Lett. 41 (2024) 017401. https://doi.org/10.1088/0256-307X/41/1/017401.

[42] Y. Zhu, D. Peng, E. Zhang, B. Pan, X. Chen, L. Chen, H. Ren, F. Liu, Y. Hao, N. Li, Z. Xing, F. Lan, J. Han, J. Wang, D. Jia, H. Wo, Y. Gu, Y. Gu, L. Ji, W. Wang, H. Gou, Y. Shen, T. Ying, X. Chen, W. Yang, H. Cao, C. Zheng, Q. Zeng, J. Guo, J. Zhao, Superconductivity in pressurized trilayer $La_4Ni_3O_{10-\delta}$ single crystals, Nature 631 (2024) 531–536. https://doi.org/10.1038/s41586-024-07553-3.

[43] M. Zhang, C. Pei, D. Peng, X. Du, W. Hu, Y. Cao, Q. Wang, J. Wu, Y. Li, H. Liu, C. Wen, J. Song, Y. Zhao, C. Li, W. Cao, S. Zhu, Q. Zhang, N. Yu, P. Cheng, L. Zhang, Z. Li, J. Zhao, Y. Chen, C. Jin, H. Guo, C. Wu, F. Yang, Q. Zeng, S. Yan, L. Yang, Y. Qi, Superconductivity in trilayer nickelate $La_4Ni_3O_{10}$ under Pressure, Phys. Rev. X 15 (2025) 021005. https://doi.org/10.1103/PhysRevX.15.021005.

[44] X. Huang, H. Zhang, J. Li, M. Huo, J. Chen, Z. Qiu, P. Ma, C. Huang, H. Sun, M. Wang, Signature of superconductivity in pressurized trilayer-nickelate $Pr_4Ni_3O_{10-\delta}$, Chin. Phys. Lett. 41 (2024) 127403. https://doi.org/10.1088/0256-307x/41/12/127403.

[45] E. Zhang, D. Peng, Y. Zhu, L. Chen, B. Cui, X. Wang, W. Wang, Q. Zeng, J. Zhao, Bulk superconductivity in pressurized trilayer nickelate $Pr_4Ni_3O_{10}$ single crystals, Phys. Rev. X 15 (2025) 021008. https://doi.org/10.1103/PhysRevX.15.021008.

[46] M. Shi, Y. Li, Y. Wang, D. Peng, S. Yang, H. Li, K. Fan, K. Jiang, J. He, Q. Zeng, D. Song, B. Ge, Z. Xiang, Z. Wang, J. Ying, T. Wu, X. Chen, Absence of superconductivity and density-wave transition in ambient-pressure tetragonal $La_4Ni_3O_{10}$, Nat. Commun. 16 (2025) 2887. https://doi.org/10.1038/s41467-025-57264-0.

[47] Z. Zhang, M. Greenblatt, Synthesis, structure, and properties of $Ln_4Ni_3O_{10-\delta}$ (Ln = La, Pr, and Nd), J. Solid State Chem. 117 (1995) 236–246. https://doi.org/10.1006/jssc.1995.1269.

[48] S. Huangfu, X. Zhang, A. Schilling, Correlation between the tolerance factor and phase transition in $A_{4-x}B_xNi_3O_{10}$ (A and B = La, Pr, and Nd; x = 0, 1, 2, and 3), Phys. Rev. Res. 2 (2020) 033247. https://doi.org/10.1103/PhysRevResearch.2.033247.

[49] D. Rout, S.R. Mudi, M. Hoffmann, S. Spachmann, R. Klingeler, S. Singh, Structural and physical properties of trilayer nickelates $R_4Ni_3O_{10}$ (R = La, Pr, and Nd), Phys. Rev. B 102 (2020) 195144. https://doi.org/10.1103/PhysRevB.102.195144.

[50] P.D. Battle, M.A. Green, J. Lago, J.E. Millburn, M.J. Rosseinsky, J.F. Vente, Crystal and magnetic structures of $Ca_4Mn_3O_{10}$, an n=3 Ruddlesden−Popper compound, Chem. Mater. 10 (1998) 658–664. https://doi.org/10.1021/cm970647r.

[51] V.M. Goldschmidt, Die gesetze der krystallochemie, Naturwissenschaften 14 (1926) 477–485. https://doi.org/10.1007/BF01507527.

[52] Y.Q. Jia, Crystal radii and effective ionic radii of the rare earth ions, J. Solid State Chem. 95 (1991) 184–187. https://doi.org/10.1016/0022-4596(91)90388-X.

[53] R.D. Shannon, Revised effective ionic radii and systematic studies of interatomic distances in halides and chalcogenides, Acta Crystsllogr. A 32 (1976) 751–767. https://doi.org/10.1107/S0567739476001551.

[54] A. Olafsen, H. Fjellvåg, B.C. Hauback, Crystal structure and properties of $Nd_4Co_3O_{10+\delta}$ and $Nd_4Ni_3O_{10-\delta}$, J. Solid State Chem. 151 (2000) 46–55. https://doi.org/10.1006/jssc.2000.8620.

[55] M.R. Norman, Landau theory of the density wave transition in trilayer Ruddlesden-Popper nickelates, Phys. Rev. B 112 (2025) 075149. https://doi.org/10.1103/43qs-x65n

[56] Y. Zhou, J. Guo, S. Cai, H. Sun, C. Li, J. Zhao, P. Wang, J. Han, X. Chen, Y. Chen, Q. Wu, Y. Ding, T. Xiang, H. Mao, L. Sun, Investigations of key issues on the reproducibility of high-$T_c$ superconductivity emerging from compressed $La_3Ni_2O_7$, Matter Radiat. Extremes 10 (2025) 027801. https://doi.org/10.1063/5.0247684.

[57] J. Li, D. Peng, P. Ma, H. Zhang, Z. Xing, X. Huang, C. Huang, M. Huo, D. Hu, Z. Dong, X. Chen, T. Xie, H. Dong, H. Sun, Q. Zeng, H. Mao, M. Wang, Identification of superconductivity in bilayer nickelate $La_3Ni_2O_7$ under high pressure up to 100 GPa, Natl. Sci. Rev. (2025) nwaf220. https://doi.org/10.1093/nsr/nwaf220.

[58] L. Wang, Y. Li, S.-Y. Xie, F. Liu, H. Sun, C. Huang, Y. Gao, T. Nakagawa, B. Fu, B. Dong, Z. Cao, R. Yu, S.I. Kawaguchi, H. Kadobayashi, M. Wang, C. Jin, H. Mao, H. Liu, Structure responsible for the superconducting state in $La_3Ni_2O_7$ at high-pressure and low-temperature conditions, J. Am. Chem. Soc. 146 (2024) 7506–7514. https://doi.org/10.1021/jacs.3c13094.

[59] J. Li, C.-Q. Chen, C. Huang, Y. Han, M. Huo, X. Huang, P. Ma, Z. Qiu, J. Chen, X. Hu, L. Chen, T. Xie, B. Shen, H. Sun, D.-X. Yao, M. Wang, Structural transition, electric transport, and electronic structures in the compressed trilayer nickelate $La_4Ni_3O_{10}$, Sci. China Phys. Mech. Astron. 67 (2024) 117403. https://doi.org/10.1007/s11433-023-2329-x.

[60] G.M. Sheldrick, A short history of SHELX, Acta Crystallogr. A 64 (2008) 112–122. https://doi.org/10.1107/S0108767307043930.

[61] O.V. Dolomanov, L.J. Bourhis, R.J. Gildea, J.A.K. Howard, H. Puschmann, OLEX2: a complete structure solution, refinement and analysis program, J. Appl. Crystallogr. 42 (2009) 339–341. https://doi.org/10.1107/S0021889808042726.

[62] Rigaku Oxford Diffraction, (2024), CrysAlis$^{Pro}$ Software system, version 1.171.43.137a, Rigaku Corporation, Wroclaw, Poland.

[63] G.M. Sheldrick, SHELXT–Integrated space-group and crystal-structure determination, Acta Crystallogr. Sec. A: Found. Adv. 71 (2015) 3–8. https://doi.org/10.1107/S2053273314026370.

[64] H.M. Rietveld, A profile refinement method for nuclear and magnetic structures, J Appl. Crystallogr. 2 (1969) 65–71. https://doi.org/10.1107/S0021889869006558.

[65] G. Kresse, J. Furthmüller, Efficiency of ab-initio total energy calculations for metals and semiconductors using a plane-wave basis set, Comput. Mater. Sci. 6 (1996) 15–50. https://doi.org/10.1016/0927-0256(96)00008-0.

[66] G. Kresse, J. Furthmüller, Efficient iterative schemes for ab initio total-energy calculations using a plane-wave basis set, Phys. Rev. B 54 (1996) 11169–11186. https://doi.org/10.1103/PhysRevB.54.11169.

[67] G. Kresse, D. Joubert, From ultrasoft pseudopotentials to the projector augmented-wave method, Phys. Rev. B 59 (1999) 1758–1775. https://doi.org/10.1103/PhysRevB.59.1758.

[68] G. Kresse, J. Hafner, Ab initio molecular dynamics for liquid metals, Phys. Rev. B 47 (1993) 558–561. https://doi.org/10.1103/PhysRevB.47.558.

[69] J.P. Perdew, K. Burke, M. Ernzerhof, Generalized gradient approximation made simple, Phys. Rev. Lett. 77 (1996) 3865–3868. https://doi.org/10.1103/PhysRevLett.77.3865.